\documentclass[conference]{IEEEtran}

\ifCLASSINFOpdf
\else
\fi
  \usepackage{amssymb}
\usepackage{latexsym}
\usepackage{amsfonts}
\usepackage{amsthm}
\usepackage{amsmath}
\usepackage{graphics}
\usepackage{setspace}
\usepackage{graphicx}
\usepackage{epstopdf}
\usepackage{color}
\usepackage{float}
\usepackage{wrapfig}
\usepackage{color}
\usepackage{rotating}
\usepackage{array} 
\usepackage{dblfloatfix}
\usepackage{breakurl}
\usepackage[hyphens]{url}

\usepackage[table]{xcolor}
\usepackage{multirow}

\usepackage{xcolor}
\usepackage{graphicx}

\definecolor{litered}{RGB}{246,194,197}
\definecolor{liteyellow}{RGB}{250,227,199}
\definecolor{litegreen}{RGB}{206,235,183}

\newcommand{\llr}{\cellcolor{litered}} 
\newcommand{\lly}{\cellcolor{liteyellow}} 
\newcommand{\llg}{\cellcolor{litegreen}}

\begin{document}

\title{Exploring the Suitability of BLE Beacons to Track Poacher Vehicles in Harsh Jungle Terrains}

\author{\IEEEauthorblockN{Karan Juj}
\IEEEauthorblockA{Cardiff University, United Kingdom\\
JujK@cardiff.ac.uk}
\and
\IEEEauthorblockN{Charith Perera}
\IEEEauthorblockA{Cardiff University, United Kingdom\\
charith.perera@ieee.org}}

\maketitle

\begin{abstract}
Our overall aim is focused on exploring whether we could use Bluetooth Low Energy (BLE) technology to track poacher vehicles in remote and rural areas such as Sabah, in Malaysia, specially deep inside the jungle terrain with little or no communication technologies exists. Tracking technologies are currently limited to relying on satellites or cellular towers, for environments that do not permit access to these signals, very few viable alternatives exist. This paper explores the use of BLE as a method to track vehicles. It works by mounting Bluetooth beacons beside a road and placing a receiver concealed somewhere inside the vehicle. As the vehicle drives past the beacon, the receiver and beacon are momentarily in range, the receiver then stores a unique ID from the beacon and when the vehicle is then in an area with GSM signal, an SMS is sent containing the unique IDs of the beacons that have been detected. This project is prototyped and tested in collaboration the Danau Girang Field Centre in Sabah, Malaysia. The results offer insights for how effective BLE beacons are in a tracking situation for where the beacon and receiver are in range for a short period of time as well as how different obstructions will affect the range and strength of the signal. It is important to note that our objective is not to catch the poacher, instead to understand how they move around within jungle terrain, as we can use such information to develop a comprehensive plan against poaching activities. 

\end{abstract}

\IEEEpeerreviewmaketitle

\section{Introduction}
The decline of the natural world has become one of the biggest talking points in the global news in the past decade. Poaching is one of largest impacting reasons for affecting not only the animals that are being poached but the entire surrounding ecosystem. These poachers operate in outlawed organised gangs as it is an extremely profitable industry. These organised gangs are extremely well structured, and thus difficult to track. Due to the limited technological options, it is currently almost impossible to track the poachers without them knowing once they enter the jungle. In this paper, we build a solution using Bluetooth beacons situated around the jungle and a receiver placed discretely inside the vehicle. The receiver will be mounted unbeknown to the poachers, allowing them to be tracked as they pass certain locations. We deployed and test our solution in the Danau Girang Field Centre in Sabah, Malaysia by allowing the wildlife officers the means to track poachers to further understand how they operate within the inner jungles.  

A BLE based location and tracking system also provides an alternative method to the limited technologies currently available on the market, this method  allow tracking in remote environments where other technologies have little to no signal available. Our approach would also be suitable to be utilised in other locations where other tracking methods are unsuitable.  \textbf{Contribution:}  In this paper, we present a real world study  conducted in a Malaysian jungle terrain to explore how could we use BLE technology to track poaching vehicles in harsh conditions such as high humidity and dense trees.

\section{Problem Formulation}

\subsection{Context}

Poaching is often the most lucrative industry in a local environment. The average salary in Sabah is 1,240 RYM  which is 1.98 times lower than the average national salary, those without any formal education are earning an average of 1000 RYM \cite{TheEdgeMarkets2018}. Poaching gives those in the lower societal classes an opportunity to earn significantly more money than traditional employment options. The effect of poaching on a local environment is huge as it has been shown to cause entire species to become extinct and in turn disrupt the entire ecosystems. This is an important demonstration of why poaching is significant as this effect in the jungles of Sabah would be devastating, to the local environment. If poaching was to escalate in the jungles of Sabah it would have disastrous national effects.

This project  implemented in the Lower Kinabatangan Wildlife Sanctuary in Sabah, Malaysia and used as a tool for the Danau Girang Field Centre (DGFC) to prototype to track these poachers to gain a further understanding of how the networks of poaching gangs operate. The main environment the poaching happens is where animal life is densest, in the deep jungle. This jungle surrounds the huge Kinabatangan river which runs for 560km and has a basin area of 16,800km. The roads which the poachers drive within the jungle are underdeveloped and are made of mud, therefore they will be uneven causing them to drive at low speeds which allow the solution more time to detect them.

\subsection{Requirement Specification}

Primary objective is to track vehicles (i.e., poacher vehicles) that moves deep within the jungles terrain. Through preliminary investigation and experience from locals, we learnt that most of the common technologies used to track vehicles in urban areas are  infeasible to deploy at scale in this particular area (i.e., Kinabatangan Wildlife Sanctuary) due to lack of signal penetration, communication infrastructure, and higher amount of obstacles (e.g., humidity, jungle). Further, an ideal tracking system should operate for a long time without requiring to replacement the energy sources (e.g., batteries). Due to low energy consumption and low cost, we hypothesis that BLE beacons would help us to track vehicles in this area. A brief summary of available technology is presented in Table \ref{Comparison}.

\begin{table}[h!]
	\centering
	\footnotesize
		\caption{Comparison of candidate Technologies \cite{zou2017accurate, mainetti2014survey, collotta2018bluetooth, Cao2009}}
	\begin{tabular}{|l|l|l|l|l|}
		\hline
		Technology     & Range                          & Cost   & Power  &  Environment \\ \hline \hline
		GPS            & Global                         & Medium & High              & Outdoor              \\ 
		GSM            & 45 miles& Low    & Medium            & Indoor \ Outdoor   \\ 
		Infared        & 1-5m                            & Low    & Medium            & Indoor               \\ 
		Acustic Signal & 2-10m                           & Low    & Medium            & Indoor               \\ 
		RFID           & 1-10m                           & Low    & Low               & Indoor               \\ 
		WIFI           & 20-59m                          & Low    & High              & Indoor               \\ 
		Bluetooth      & 1-30m                           & Low    & Medium            & Indoor \ Outdoor   \\ 
		BLE            & 1-100m                          & Low    & Low               & Indoor / Outdoor   \\ \hline
	\end{tabular}%

	\label{Comparison}
\end{table}

\section{System Architecture}

Our proposed BLE beacon based tracking system comprises of three components:
\begin{itemize}
	\item A receiver which is placed on a poachers vehicle
	\item A BLE beacon which will be strategically placed beside a road within the jungle terrain
	\item A software platform to log where poachers vehicles have been spotted.
\end{itemize}

\subsection{Receiver}

The receiver  will contain BLE functionality which will act as the observer which will be 'listening' out for the beacons signal. The receiver will also contain SMS capabilities which would be used to send SMS messages to a SMS server to plot on a map which beacons have been spotted. The information on the text messages regarding the beacons that have been detected will correlate to the locations where the poachers have been detected.

The approach is to use an informant to place the receiver on a specific discrete part of the vehicle.. The poachers will then continue their journey unaware that there is a receiver on their vehicle. When the vehicle passes one of the BLE beacons the receiver will pick up the unique name of the device. The vehicle will then continue their journey passing more beacons which the receiver should detect and log. When the vehicle leaves the jungle to go to the city to sell the poached animals, the GSM module will then send an SMS containing the unique names of the beacons which have been detected, to the SMS server.

\subsection{Bluetooth Beacons}
The BLE beacons will be placed at specific locations on the road. The beacon will display a unique identifier so that it cannot be confused which beacon has been seen by the receiver. Due to the possibility of the receiver travelling at speeds which could cause the beacon to not be detected the beacons will be placed specifically in locations where a vehicle would have to drive extra slow such as a tight bend to try to maximise the amount of time the receiver has to read the beacon. In evaluation, we will investigate the best places to place a beacon such as the height of the beacon and the distance to the road \cite{pancham2018investigation}.

\subsection{Visualising the Beacon Data}
\begin{figure}[h!]	
	\centering
	\includegraphics[scale=0.35]{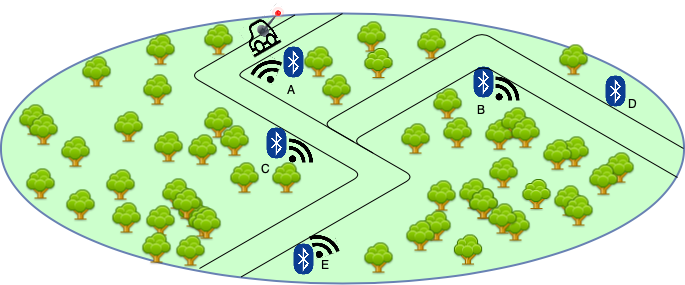}
	\caption{BLE Beacon Positioning Plan}
	\label{fig:beconlocation}
	
\end{figure}

The data that is received from the vehicle is more effective if the users can visualise where the poachers have been detected. This visualisation should allow the user to see where beacons are placed within the jungle so when the user receives the SMS they can see geographically the points in the jungle that the vehicle has been detected. have been spotted. Figure \ref{fig:beconlocation} shows a representation of how the BLE beacons would be deployed.

\section{Evaluation}

We conducted five different experiments to evaluate different performance aspects to determine the suitability of BLE beacons for vehicle tracking.

\subsection{Performance Comparison of Bluetooth 4 and Bluetooth 5}

We selected two different beacons to test for all of the parameters, first was a HM-10 module (Bluetooth 4.0) that was powered by a LiPo battery which we had configured using AT commands to act as a beacon. The second beacon was an off-the-shelf module (Bluetooth 5.0). To test the range of the beacons,  first test that we conducted was designed to test the range of the beacons. The test works by placing a beacon in a fixed position with varying obstructions of different materials to simulate both potential casing materials and objects which could affect the performance of the signal between the beacon and the receiving device in the jungle. We took measurements of the receiver testing the RSSI every two meters. RSSI stands for Received Signal Strength Indicator, it is a measurement of how well the receiver can hear the signal from an access point or router it is useful as it can tell us how good of a connection the receiver will get to the beacon at any given point. Both beacons RSSI are meant to be set at -70dBm at a 1-meter range from the beacon. The first test was to test the range of both beacons. The test works by placing a beacon in a fixed position with no obstructions between the beacon and the receiving device.

\begin{figure}[h!]
	\centering
	\includegraphics[scale=0.32]{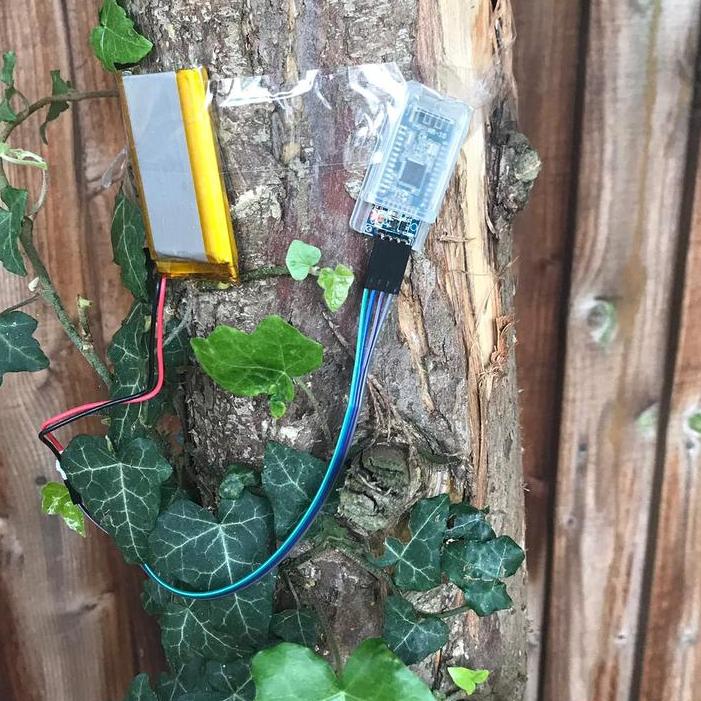}
	\caption{HM-10 beacon mounted on a tree}
	\label{fig:HM10NoObstruction}
\end{figure}

\begin{figure}[h!]
	\centering
	\includegraphics[scale=0.30]{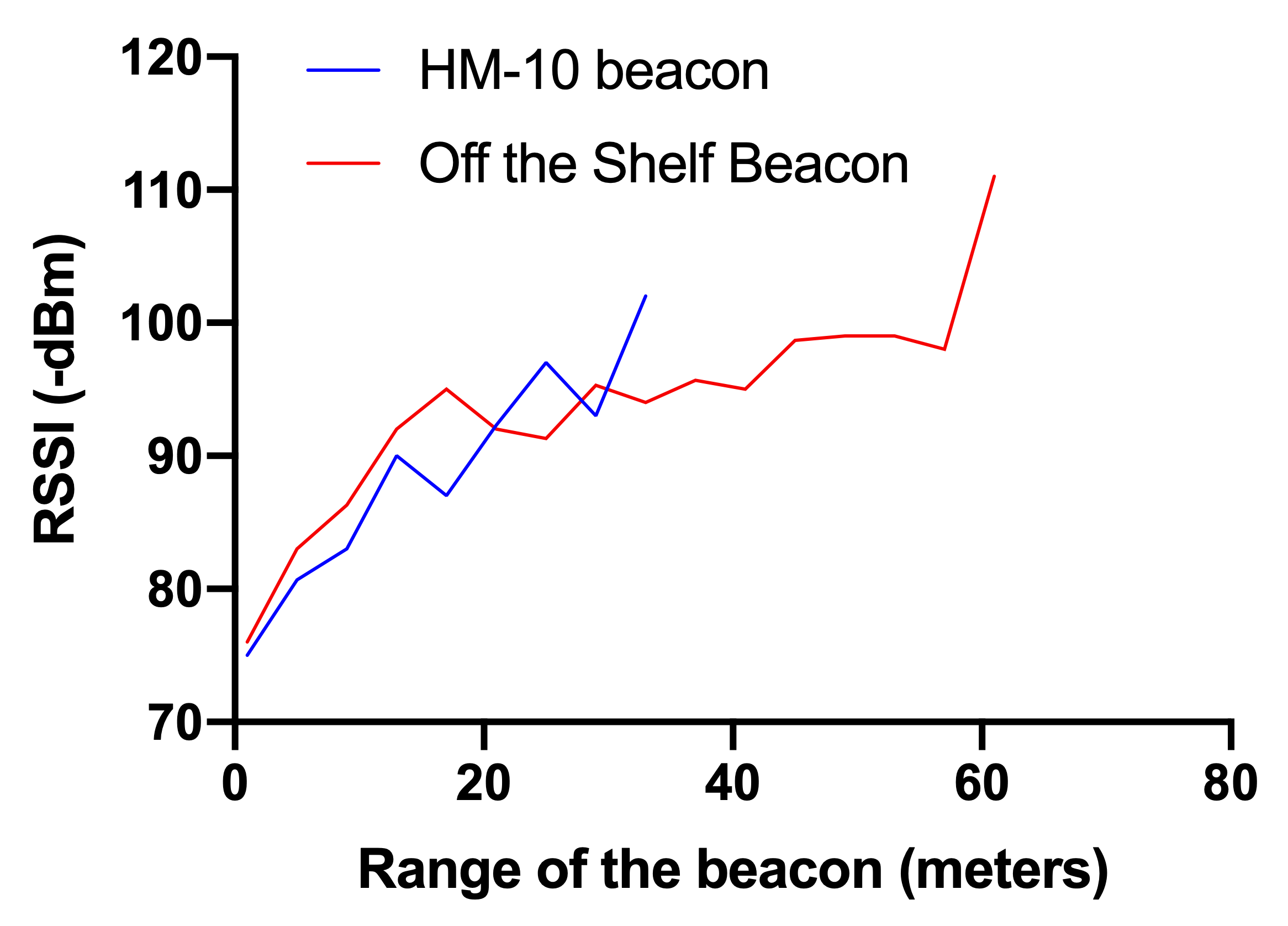}
	\caption{RSSI from the beacons at increasing distances}
	\label{fig:BeaconRangeGraph}
\end{figure}

We tested in a large field where no other devices could be detected and tested one beacon at a time to ensure no interference this is important to isolate the results of the test to ensure a fair result. We averaged the results and presented in Figure \ref{fig:BeaconRangeGraph} (RSSI on the Y-axis, the lower the RSSI the stronger the signal). On the X-axis we see the range of the beacons. According to our results, the range of the off-the-shelf beacon is 84\% larger, and throughout that range, the signal strength remains to be strong until 41 meters as RSSI begins to deteriorate in reliability from the beacon after 95 RSSI which the HM-10 beacon reached at 25 meters. This shows that the off the shelf beacon has a significantly longer range while maintaining a good signal.This result was expected  as the HM-10 uses Bluetooth 4.0 where the off-the-shelf beacon uses Bluetooth 5.0 which has been developed for an increased range. 

\subsection{Effects of Obstructions of Different Materials on the Beacon Signal}

\begin{figure}[h!]
	\centering
	\includegraphics[scale=0.76]{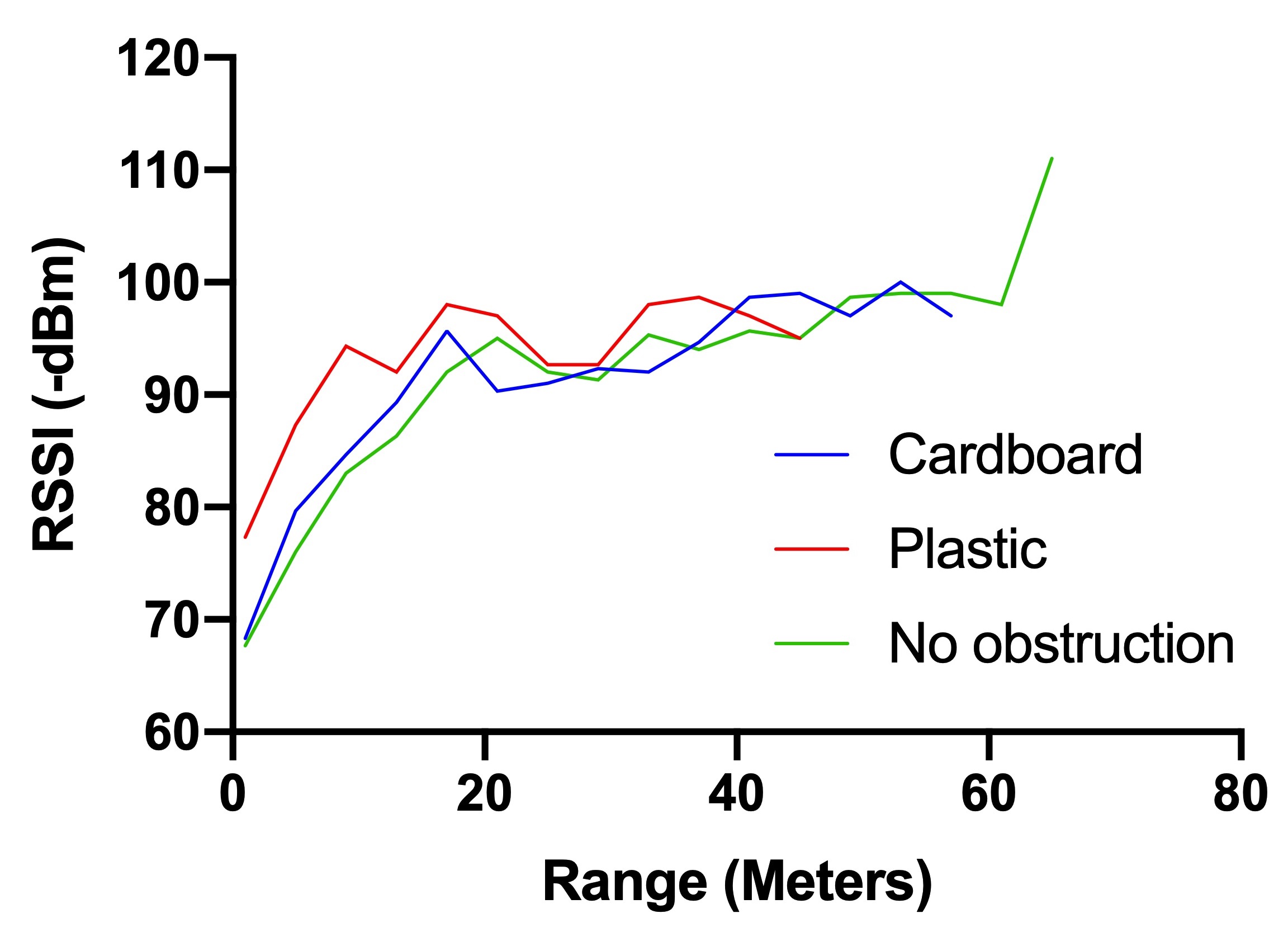}
	\caption{Affect on RSSI at distances when the beacon is obstructed by various materials}
	\label{fig:CardboardPlasticNoObst}
\end{figure}

\begin{figure*}[h!]
	\centering
	\includegraphics[scale=0.38]{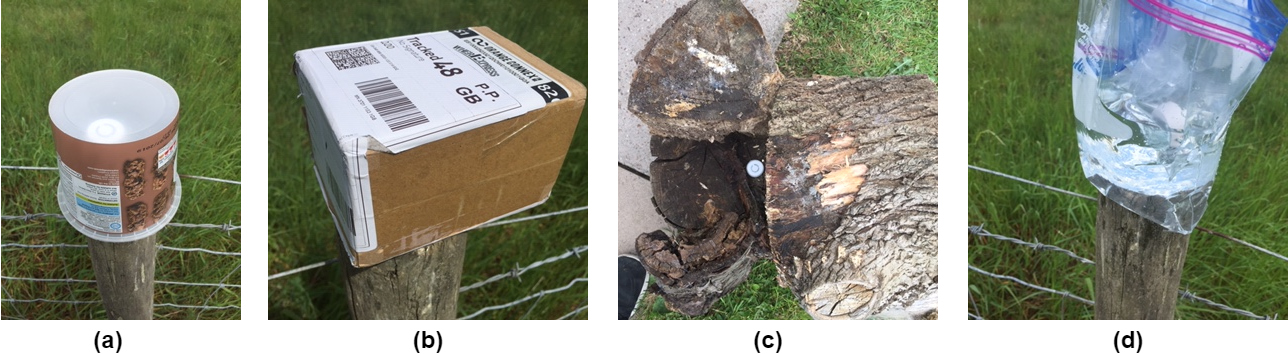}
	\caption{Different BLE beacon positionings}
	\label{fig:DifferentLocations}
\end{figure*}

The next test aimed to test the various types of obstructions that the signal would face in deployment. As shown in Figure \ref{fig:DifferentLocations}, we tested plastic and cardboard cases which were intended to simulate the effect that adding a case would have on the signal. We  tested the RSSI at distances increasing by four meters starting at one meter. The results are presented in Figure \ref{fig:CardboardPlasticNoObst}. The graph contains the data from the unobstructed signal to allow for easier comparison. We can see that for the plastic case the RSSI is higher at one meter than in the unobstructed and the cardboard, from this we can tell instantly that plastic affects the signal strength significantly. Towards the 25 meter range, the effect of the plastic seems to level out as all of the results were very similar at this range. The case materials did show to affect the range however with the plastic losing signal at 45 meters resulting in a loss of range versus the unobstructed signal of 44\%  and the cardboard losing signal at 57 meters resulting in a loss of 14\%. This leaves an interesting trade-off for the casing as a plastic casing would have been the ideal material as it is waterproof and discrete.

\subsection{Factors that Effect the Signal in Jungle Terrains}
The next test we did was to see how the signal would be affected within a jungle environment. Sabah receives 2500-3500 mm of rainfall annually, for comparison Cardiff receives 991 mm of rain annually. It is also up to 100\% humidity in the jungle so it is important to know how water would effect the signal strength. This proved to be a difficult test as any container that held the water would add to the obstruction of the signal which would make it difficult to test the affect of water without the container. For this, we chose a sandwich bag as it is extremely thin plastic which would have the least effect on the signal as possible. It would be naive to say that the obstruction from the sandwich bag did not effect the results on the signal at all, for this reason before testing the water we put the beacon in sandwich bags and tested this individually to understand the affect of the plastic bags on the signal so we can better understand the effect of the water. 

We filled this with one litre of water and submerged the beacon to try and replicate the wetness of the surroundings in the jungle. The graph below shows the results of the data. For the `just water' data as we had the data for the water and plastic bag combined, and for the plastic bag we took the difference of the data for the unobstructed signal and added it on to the water data to try and get a gauge for how water alone would affect the signal. The graph below shows that the plastic bag had a small impact but the water itself had the biggest impact on the signal of any material we have tested having a range of only 33 meters. From the trends that we have seen in the previous graphs, based on the results of the just water we could expect another 4 meters of range based on the RSSI strength of 29 and 33 meters, bringing the range of the water up to 37 meters.


\begin{figure}[h!]
	\centering
	\includegraphics[scale=0.30]{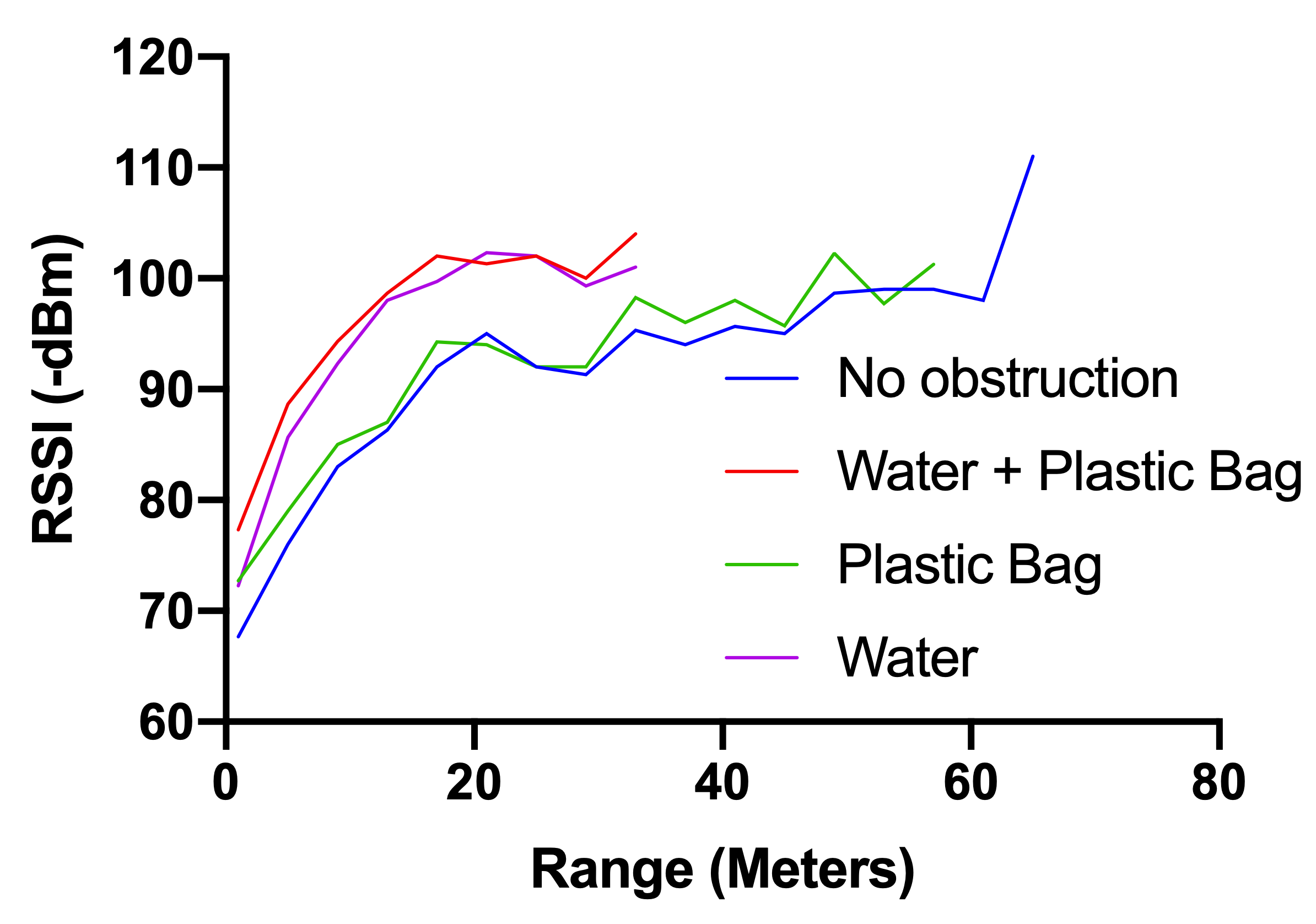}
	\caption{Affect on RSSI at distances when the beacon is obstructed by water}
	\label{fig:EffectOfWater}
\end{figure}

\subsection{Effects of Broadcast Interval and Vehicle Speed}

The driving test is the next experiment which was needed. As the beacon is to be placed in the jungle onto a tree or similar surface, we mounted the beacon on to a tree similar to how a beacon would be mounted in the jungle. Figure \ref{fig:BeaconOnRoad} shows the first road which we tested on, we tried to test in the closest environment to the jungle as possible, this road is in Watford, England. The road is pictured below and the mounted beacon is identified. In this experiment, we are driving at various velocities to see whether the receiver picks up the signal from the beacon. The estimated max velocity was 70 Kph which is approximately 43.4 mph so testing was planned to go up to 45 mph although it was not expected to be successful at the higher speeds. For each speed, the broadcast interval would be increased to test what the highest broadcast interval we could set the beacon on while getting reliable results to ensure the beacon is detected every time that the vehicle is in range. The test recorded whether when the vehicle drove past the beacon it was detected or not. 

\begin{figure}[h!]
	\centering
	\includegraphics[scale=0.128]{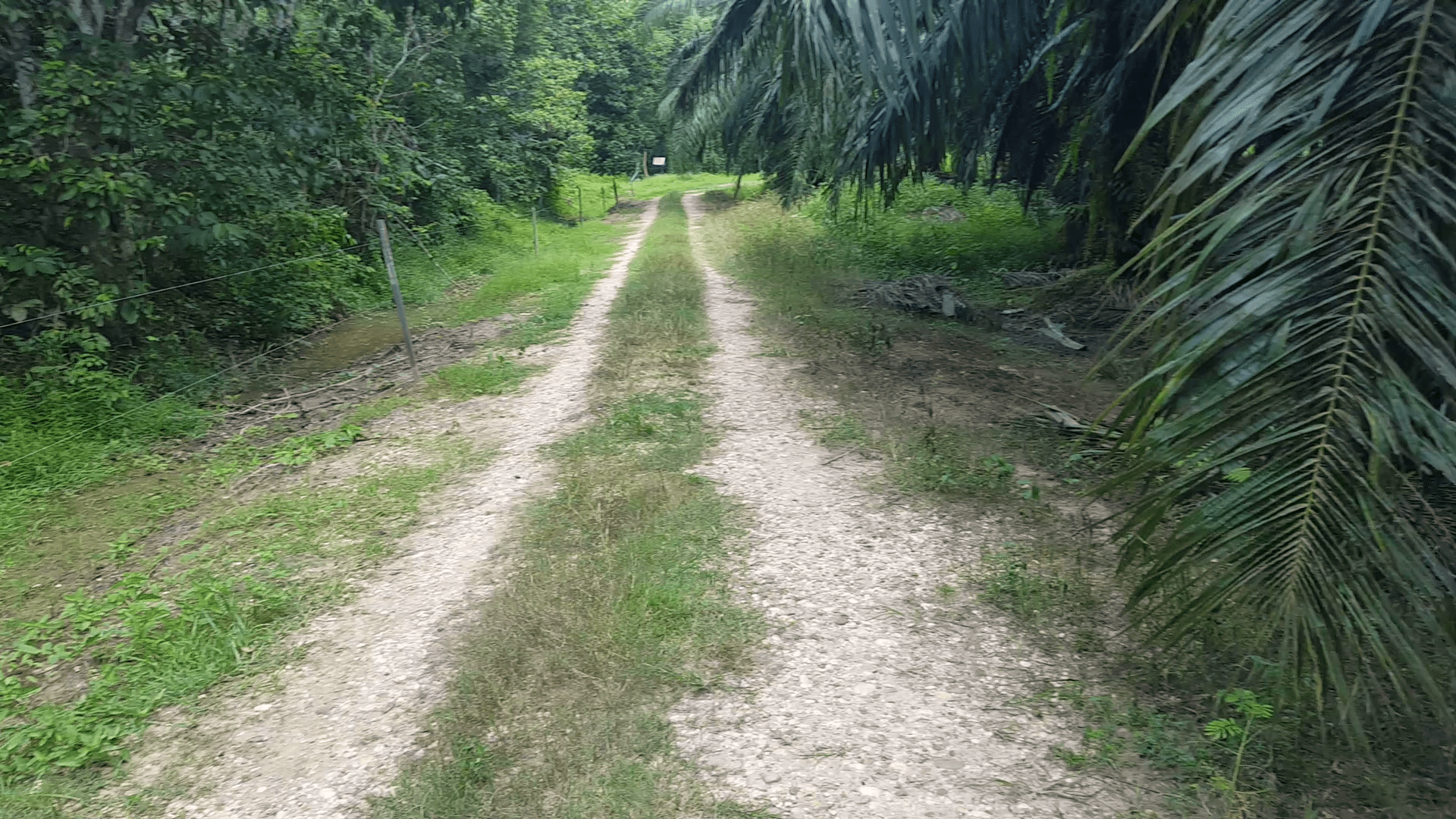}
	\caption{A road in Sabah where poachers move around (Buffer area in between sanctuary and palm plantations fields)}
	\label{fig:Road}
\end{figure}

This road which had a similar build to go a jungle road was difficult to drive on over 20 mph so on this road we would test at 20 mph to maintain safety, if the beacon was not detected, then we would decrease speed to see what the maximum speed the beacon was detected at that broadcast interval. We started the broadcast interval at 200ms and the receiver picked the beacon up every time. We repeated this test for broadcast intervals up to 1600ms, increasing in 100 intervals. The receiver detected the beacon every time without fail. Due to the success of this test, a tarmac road where a vehicle could get to higher speeds was needed. To test with higher speeds, we used better roads outside jungle terrain that allows us to drive with higher speeds safely. The closest the vehicle was to the beacon at its peak was approximately two meters away. 

We conducted the same experiment as the previous experiment on the dirt road but with the difference that the vehicle would now be travelling up to 45 mph. This worked with flawlessly at 45mph until the beacon interval was increased to 600ms, with the lower beacon intervals the receiver had detected the beacon before passing it. As it went to 800ms it was detected but about half a second after passing the beacon each time. 1000ms was also detected far after the beacon but now about 1.5 seconds after passing the beacon. The 1200ms broadcast interval is where the receiver started to not detect the beacon. As the test was repeated 3 times the beacon was detected 2 out of the 3 times tested, one of the times tested the beacon was detected approximately 25 meters after passing the beacon. This is due to the probability of beacon advertising and the HM-10 scanning while the vehicle is in range. The probability that detection will occur decreases exponentially as the broadcast interval increases.

\begin{table}[]
	\centering
	\scriptsize
		\caption{Effect of Different Broadcast Intervals and Speed of the Vehicles}
	\begin{tabular}{|l|l|l|l|l|l|l|l|}
		\hline
		& \rotatebox{90}{1000ms} & \rotatebox{90}{1100ms} & \rotatebox{90}{1200ms} & \rotatebox{90}{1300ms} & \rotatebox{90}{1400ms} & \rotatebox{90}{1500ms} & \rotatebox{90}{1600ms} \\ \hline
		5 mph  & Y \llg      & Y  \llg    & Y  \llg    & Y  \llg    & Y  \llg    & Y \llg     & Y \llg     \\ 
		10 mph & Y \llg      & Y \llg     & Y \llg     & Y \llg     & Y \llg     & Y  \llg    & Y \llg     \\ 
		15 mph & Y \llg     & Y \llg     & Y \llg     & Y \llg     & Y \llg     & Y  \llg    & Y  \llg    \\ 
		20 mph & Y \llg     & Y \llg     & Y \llg     & Y \llg     & Y  \llg    & Y \llg     & Y  \llg    \\ 
		25 mph & Y \llg     & Y \llg     & Y \llg     & Y  \llg    & Y \llg     & 66\% \lly   & 66\% \lly   \\ 
		30 mph & Y \llg     & Y \llg     & Y \llg     & Y  \llg    & 66\% \lly   & 66\%  \lly  & 33\% \lly   \\ 
		35 mph & Y \llg     & Y \llg     & Y \llg     & Y \llg    & 66\% \lly   & 33\% \lly   & 33\% \lly   \\ 
		40 mph & Y \llg     & Y  \llg    & Y \llg     & Y  \llg    & 66\% \lly    & 33\% \lly   &  N \llr     \\ 
		45 mph & Y \llg     & Y  \llg    & 66\%  \lly   & 66\% \lly    & 33\% \lly    &  N \llr      &  N  \llr    \\ \hline
	\end{tabular}
	\\ \vspace{5pt}
	Y: the beacon was detected every time;	33\%: the beacon was detected in 33\% of the time;	66\%: the beacon was detected in 66\% of the time;	N: the beacon was not detected.

	\label{Tbl:BroadcastIntervals}
	
\end{table}

Table \ref{Tbl:BroadcastIntervals} gives a partial set of results from this test as all of the results from 200ms to 1200ms were Y's as the beacon was detected every time. The table shows when the reliability of the solution starts to deteriorate. As there are two variables in this connection the beacon and the receiver, it is important to note that the receiver is set on the fastest reliable loop cycle which is 2500ms. 


\subsection{Effects of mounted under the bonnet}

For the final test, we needed to fully test how the beacon and receiver would communicate in the jungle, due to the fact we do not want poachers to find the receiver it has to be placed somewhere hidden. We believe that the best place would be under the wheel arch. As we assume poachers will drive 4x4 vehicles the wheel arch gap will be a lot larger. However, we wanted test the variations in BLE strength by putting the BLE receiver on a place with higher obstruction.

\begin{figure}[h!]
	\centering
	\includegraphics[scale=0.35]{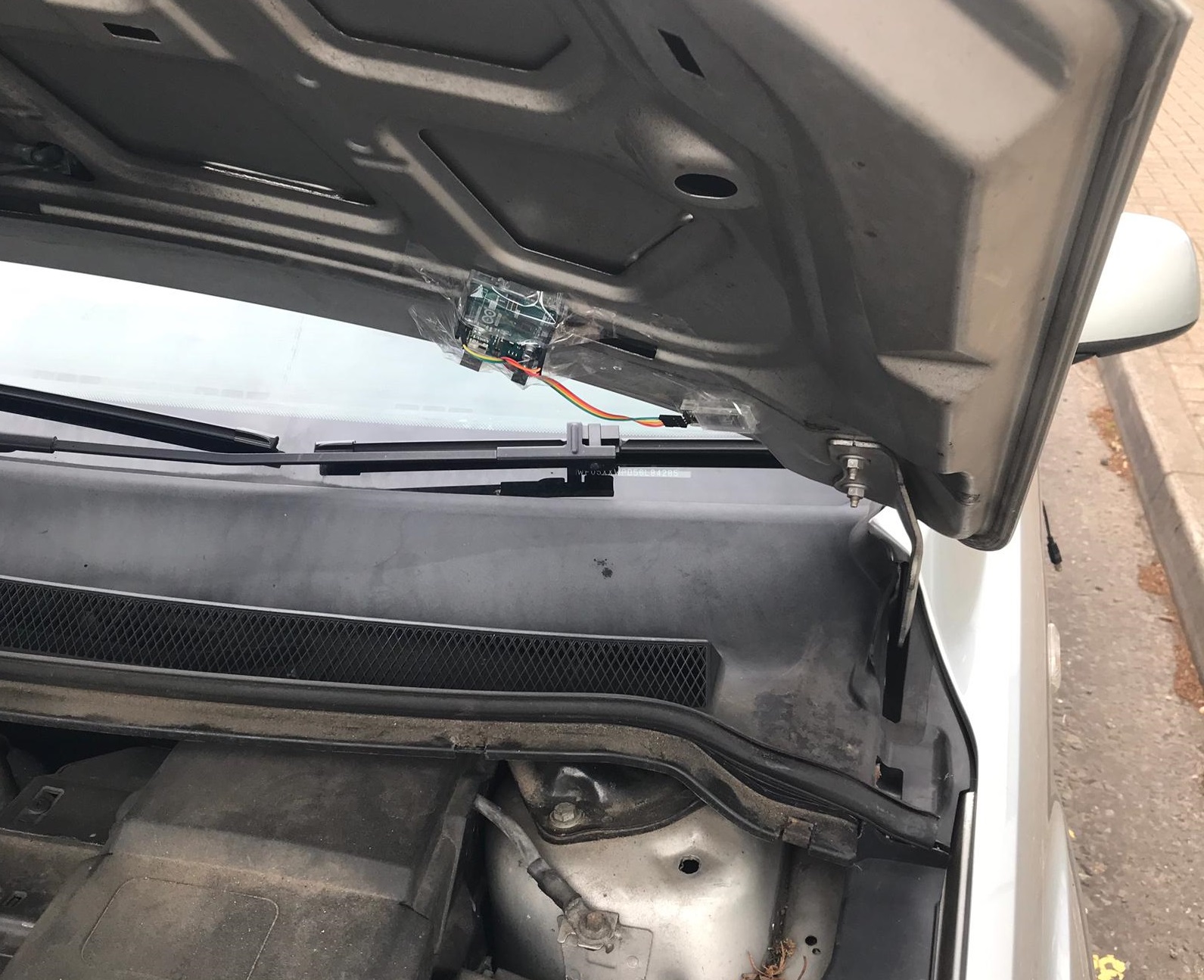}
	\caption{Deployment of the BLE Receiver Inside the Bonnet}
	\label{fig:Receiveronbonnet}
\end{figure}

Therefore, we fastened the receiver to the bonnet as shown in Figure \ref{fig:Receiveronbonnet}. A bonnet will have more of an impact on the signal as it is surrounded by far more and thicker metal being that close to the engine, this also would not be a feasible location to put the receiver in deployment as the heat of the engine could potentially damage the receiver.

Results are presented in Table \ref{Tbl:BroadcastIntervalsBonnet}, the results from 200ms to 700ms were all Y's which means that the receiver detected the beacon on each of the 3 tests. The results were significant showing the huge impact that concealing the receiver has on the signal strength. The results at 800ms were similar to the results of the previous test at 1200ms and the results at 1000ms in this test were similar to that of 1400 in the previous test. This shows a result that with the receiver concealed the signal is approximately as effective as if the beacon was set to -400ms less than any broadcast interval with it concealed.

\begin{table}[]
	\centering
		\caption{Effect of Different Broadcast Intervals and Speed of the Vehicles - Receiver Mounted Under the Bonnet}
	\scriptsize
		\begin{tabular}{|l|l|l|l|l|l|l|l|l|l|}
			\hline
			 & \rotatebox{90}{700ms} & \rotatebox{90}{800ms} & \rotatebox{90}{900ms} & \rotatebox{90}{1000ms} & \rotatebox{90}{1100ms} & \rotatebox{90}{1200ms} & \rotatebox{90}{1300ms} & \rotatebox{90}{1400ms} & \rotatebox{90}{1500ms}  \\ \hline
			\textbf{5 mph}                & Y   \llg           & Y  \llg            & Y     \llg         & Y  \llg             & Y    \llg           & Y  \llg             & Y  \llg             & Y   \llg            & N    \llr                    \\ 
			\textbf{10 mph}         & Y   \llg           & Y  \llg            & Y  \llg            & Y   \llg            & Y  \llg             & Y  \llg             & Y  \llg             & 66\%   \lly          & N  \llr                        \\ 
			\textbf{15 mph}       & Y  \llg            & Y  \llg            & Y   \llg           & Y  \llg             & Y  \llg             & Y  \llg             & Y   \llg            & 66\%  \lly           & N   \llr                       \\ 
			\textbf{20 mph}         & Y  \llg            & Y    \llg          & Y  \llg            & Y  \llg             & Y  \llg             & Y  \llg             & 66\%  \lly           & 33\%  \lly           & N    \llr                     \\ 
			\textbf{25 mph}         & Y   \llg           & Y  \llg            & Y  \llg            & Y   \llg            & Y  \llg             & Y  \llg             & 66\%   \lly          & 33\%   \lly          & N   \llr                         \\ 
			\textbf{30 mph}               & Y   \llg           & Y  \llg            & Y   \llg           & Y     \llg          & 66\%   \lly          & 66\%   \lly          & 33\%  \lly           & N    \llr         & N   \llr                       \\ 
			\textbf{35 mph}               & Y  \llg            & Y  \llg            & Y   \llg           & 66\%     \lly        & 33\%  \lly           & 66\%   \lly          & N   \llr            & N      \llr         & \llr N                           \\ 
			\textbf{40 mph}               & Y   \llg           & 66\%   \lly         & 66\%  \lly            & 66\%  \lly           & 66\%  \lly           & 66\%  \lly           & N \llr              & N  \llr              &         N \llr       \\
			\textbf{45 mph}             & Y  \llg            & 66\%  \lly          & 66\%  \lly          & 33\%  \lly           & 33\%  \lly           & N    \llr            & N    \llr            & N      \llr          &    N \llr        \\     \hline
		\end{tabular}%
	\\ \vspace{5pt}
Y: the beacon was detected every time;	33\%: the beacon was detected in 33\% of the time;	66\%: the beacon was detected in 66\% of the time;	N: the beacon was not detected.
\vspace{-5pt}
	\label{Tbl:BroadcastIntervalsBonnet}
\end{table}

\section{Discussion, Lessons Learnt, and Recommendation}

Following the results of the receiver under the bonnet test, it shows that under the bonnet the signal strength is greatly affected. However, detection can occur with the right beacon interval. This is important because as under the bonnet is the most heavily obstructed part of the vehicle it gives us the ability to give the informant the choice of where to put the receiver on the vehicle depending on the vehicle. Placing the receiver on the side of the vehicle could have a big effect on the signal. If the receiver is placed on the left wheel arch and the beacon on a tree to the right of the road, the obstruction will be the entire width of the vehicle. As we have tested under the bonnet, we can be confident that with a low enough broadcast interval, detection would still happen every time.

To choose the broadcast interval for the beacons we need to consider the trade-off of battery life vs effectiveness. Detection every time the receiver comes in to range of the beacon is a requirement so we must choose the highest beacon interval that gives us a reliable solution and maximise battery life. The max speed would be 60-70 km/h (45 mph) as the roads are very slippery and there are a lot of bumps and holes on the dirt roads, the average speed is 40 - 50 km/h (30 mph). As 30mph is the average speed, we can immediately rule out broadcast intervals of 1100ms or higher as the results at this speed shows that detection only occurred 2 out of the 3 times it was tested. To get a reliable solution to work for the max speed a broadcast interval of 700ms would be the lowest, which would give an estimated battery life of 131.25 days.

Alternatively, when beacons are placed a smart phone can be used to use the app the beacon designers created which allows the user to connect to the beacon to change the settings such as the broadcast interval. Table \ref{Tbl:Guide} has been created  as a guide for the individuals deploying the beacons if they have smart phones available in the jungle. It gives those deploying the beacons the discretion to estimate what the max speed would be for the specific road which they are deploying on, based on this estimate they can see what broadcast interval they should set the beacon to and thus they can determine how long the battery would last, if they do have a smart phone and can use this method it will ensure that the battery life is maximised for each individual beacon while ensuring that the receiver will still be able to detect it. 

\begin{table}[!ht]
	\centering
	\footnotesize
	\caption{Guide to Inform How to Set Beacon Broadcast Interval  Based on the Maximum Road Speed}
		\begin{tabular}{|p{2cm}|p{2.5cm}|p{2.5cm}|}
			\hline
			Max Road Speed (mph) & Set broadcast interval to: & Estimated battery life (Days) \\ \hline
			5              & 1400ms                     & 262.5                         \\ 
			10             & 1300ms                     & 243.75                        \\ 
			15             & 1300ms                     & 243.75                        \\ 
			20             & 1200ms                     & 225                           \\ 
			25             & 1200ms                     & 225                           \\ 
			30             & 1000ms                     & 187.5                         \\ 
			35             & 900ms                      & 168.75                        \\ 
			40             & 700ms                      & 131.25                        \\ 
			45             & 700ms                      & 131.25                        \\ \hline
			
		\end{tabular}%

	\label{Tbl:Guide}
	\vspace{-10pt}
\end{table}

\section{Related Work}

Traditionally, BLE based solutions are being developed for indoor tracking. Most attractive characteristics of BLE are cheaper cost \cite{Mohebbi2017}, smaller size and long lasting battery life. For example, Altini et al. \cite{Altini2010} have used neural networks based approach to develop a indoor localization method using BLE. In another work, Molina et al. \cite{Molina2018} have used BLE to develop a indoor positioning system for airports \cite{Ndzukula2017} Further, compared to WiFi, BLE seems to work better for indoor localisation \cite{Zhao2014}. One study of using BLE in outdoor is for sightseeing. Ito et al. \cite{Ito2016} have developed a navigation system using BLE beacon for sightseeing in Nikko. There results show that BLE beacon in outdoor worked well and if visitors are walking, they can find almost all beacons. Add to this work,  we demonstrated that BLE beacons can be used to track moving vehicles.

\section{Conclusion and Future Work}
BLE has been widely used for indoor tracking. But not many attempts have been made to use BLE for outdoor tracking. Primary reason for this is that, in outdoors, most of the time, there are other technologies that work for better (specially for tracking needs). However, in this paper, addressed a outdoor tracking problem in an jungle terrain where other technologies does not work. During our study, we found that, even though deploying and BLE beacon on a poacher vehicle is challenging (without getting noticed by the poacher), it is totally feasibility to use BLE technology to tracking vehicle. Through series of studies, we evaluated and recommended how BLE beacons need to be configured and deployed (e.g., broadcasting intervals, location of the beacons). We also evaluated and recommended how and where the BLE receiver should be attached within the poachers vehicle. After extensive evaluation, we learnt that BLE beacons can be successfully used in jungle terrains to track vehicle.

\section*{Acknowledgement}

We acknowledge the support by GCRF Facilitation Funds and Cardiff University's Global Opportunity Mobility Funds.



%
\bibliographystyle{IEEEtran}
\bibliography{library}

\end{document}